\documentclass[11pt]{article}
\usepackage{aaspp4}
\newcommand\mjybm{\mbox{mJy~beam${}^{-1}$}}
\newcommand\thd{\theta_d}
\newcommand\cnsq{\ensuremath{C_n^2}}

\newcommand\smunits{\mbox{kpc~m${}^{-20/3}$}}

\lefthead{Lazio et al.}
\righthead{VLBI Images of 1741$-$038 in an ESE}

\received{1999 July~14}
\revised{1999 October~13}
\accepted{1999 October~13}
\paperid{MS~50071}
\ccc{PD}

\begin{document}
\title{The Extreme Scattering Event Toward 1741$-$038: VLBI Images}

\author{T.~Joseph~W.~Lazio}
\affil{Naval Research Laboratory, Code~7213, Washington, DC
	20375-5351, USA; lazio@rsd.nrl.navy.mil}

\author{A.~L.~Fey}
\affil{United States Naval Observatory, Code EO, 3450 Massachusetts
	Ave.~NW, Washington, DC  20392-5420, USA; afey@alf.usno.navy.mil}

\author{Brian Dennison}
\affil{Physics Department, Robeson Hall, Virginia Polytechnic
	Institute, Blacksburg, VA, 24061-0435, USA;
	dennison@astro.phys.vt.edu}

\author{F.~Mantovani}
\affil{Istituto di Radioastronomia del~CNR, Via Gobetti~101, I-40129,
	Bologna, Italy; fmantova@rossem.ira.bo.cnr.it}

\author{J.~H.~Simonetti}
\affil{Physics Department, Robeson Hall, Virginia Polytechnic
	Institute, Blacksburg, VA, 24061-0435, USA;
	jhs@astro.phys.vt.edu}

\author{Antonio Alberdi}
\affil{Instituto de~Astrof{\'\i}sica de~Andaluc{\'\i}a, CSIC,
	Apdo.~3004, Granada, Spain; alberdi@laeff.esa.es}

\author{A.~R.~Foley}
\affil{Netherlands Foundation for Research in Astronomy, Postbus~2,
	NL-7990~AA, Dwingeloo, The Netherlands; tony.foley@nfra.nl}

\author{R.~Fiedler}
\affil{Naval Research Laboratory, Code~7261, Washington, DC
	20375-5351, USA; fielder@sealab.nrl.navy.mil}

\author{M.~A.~Garrett}
\affil{Joint Institute for VLBI in Europe, Postbus~2, NL-7990~AA,
	Dwingeloo, The Netherlands; mag@jive.nfra.nl}

\author{Hisashi Hirabayashi}
\affil{Institute of Space and Astronautical Science,3-1-1 Yoshinodai,
	Sagamihara, Kanagawa 229, Japan; hirax@vsop.isas.ac.jp}
 
\author{D.~L.~Jauncey}
\affil{Australia Telescope National Facility, Epping, New South Wales
	2121, Australia; djauncey@atnf.csiro.au}

\author{K.~J.~Johnston}
\affil{US Naval Observatory, 3450 Massachusetts Avenue NW, Washington, 
	DC 20392; kjj@astro.usno.navy.mil}

\author{Jon Marcaide}
\affil{Departamento de~Astronom{\'\i}a, Universitat de~Val\`encia,
	E-46100 Burjassot, Val\`encia, Spain; jmm@vlbi.daa.uv.es}

\author{Victor Migenes}
\affil{Departamento de~Astronomia, Universidad de~Guanajuato, Apartado 
	Postal~144, 36000 Guanajuato, Gto, M\'exico;
	vmigenes@cuevano.astro.ugto.mx}
 
\author{G.~D.~Nicolson}
\affil{Hartebeesthoek Radio Astronomy Observatory,
	\hbox{P.O.}~Box~443, 1740 Krugersdorp, South Africa;
	george@bootes.hartrao.ac.za}

\and 

\author{Tiziana Venturi}
\affil{Istituto di~Radioastronomia CNR, via~Gobetti, 101, 40129
	Bologna, Italy; tventuri@ira.bo.cnr.it}

\begin{abstract}
We report multi-epoch VLBI observations of the source 1741$-$038
(OT$-068$) as it underwent an extreme scattering event.  Observations
at four epochs were obtained, and images were produced at three of
these.  One of these three epochs was when the source was near the
minimum flux density of the ESE, the other two were as the flux
density of the source was returning to its nominal value.  The fourth
epoch was at the maximum flux density during the egress from the ESE,
but the VLBI observations had too few stations to produce an image.

During the event the source consisted of a dominant, compact
component, essentially identical to the structure seen outside the
event.  However, the source's diameter increased slightly at~13~cm,
from near 0.6~mas outside of the ESE to near 1~mas during the
\hbox{ESE}.  An increase in the source's diameter is inconsistent with
a simple refractive model in which a smooth refractive lens drifted
across the line of sight to 1741$-$038.  We also see no evidence for
ESE-induced substructure within the source or the formation of
multiple images, as would occur in a strongly refractive lens.
However, a model in which the decrease in flux density during the ESE
occurs solely because of stochastic broadening within the lens
requires a larger broadening diameter during the event than is
observed.  Thus, the ESE toward 1741$-$038 involved both stochastic
broadening and refractive defocussing within the lens.  If the
structure responsible for the ESE has a size of order 1~AU, the level
of scattering within an ESE lens may be a factor of~$10^7$ larger than
that in the ambient medium.  A filamentary structure could reduce the
difference between the strength of scattering in the lens and ambient
medium, but there is no evidence for a refractively-induced elongation
of the source.  We conclude that, if ESEs arise from filamentary
structures, they occur when the filamentary structures are seen
lengthwise.

We are able to predict the amount of pulse broadening that would
result from a comparable lens passing in front of a pulsar.  The pulse
broadening would be no more than 1.1~$\mu$s, consistent with the lack
of pulse broadening detected during ESEs toward the pulsars
PSR~B1937$+$21 and PSR~J1643$-$1224.

The line of sight toward 1741$-$038 is consistent with a turbulent
origin for the structures responsible for ESEs.  The source 1741$-$038
lies near the radio Loop~I and is seen through a local minimum in
100$\mu$ emission.
\end{abstract}


\section{Introduction}\label{sec:intro}

Extreme scattering events (ESE) are a class of dramatic changes in the
flux density of radio sources (\cite{fdjws94}).  They are typically
marked by a decrease ($\gtrsim$ 50\%) in the flux density near 1~GHz
for a period of several weeks to months, bracketed by substantial
increases, viz.\ Fig.~\ref{fig:lightcurve}.  Because of the
simultaneity of the events at different wavelengths, the time scales
of the events, and light travel time arguments, ESEs are likely due to
strong scattering by the Galactic interstellar medium (ISM;
\cite{fdjh87a}; Romani, Blandford, \& Cordes~1987).  First identified
in the light curves of extragalactic sources, ESEs have since been
observed during a timing program of the pulsars PSR~B1937$+$21
(\cite{cblbadd93}; Lestrade, Rickett, \& Cognard~1998) and
PSR~J1643$-$1224 (Maitia, Lestrade, \& Cognard~1998).

Extreme scattering events are generally ascribed to be the result of
strong interstellar refraction by discrete ionized structures
(\cite{fdjh87a}; \cite{rbc87}; Clegg, Fey, \& Lazio~1998), though
Fiedler et al.~(1994a) developed a model in which ESEs are due to
stochastic broadening of source diameters.  The extent to which
refractive defocussing or stochastic broadening dominates during an
ESE may be determined through VLBI imaging of a source undergoing an
\hbox{ESE}.  If refractive defocussing dominates, the position of the
source may wander, and the shape of the lens or substructure within
the lens may produce distortions in the appearance of the source; if
the refraction is strong enough, the source may be multiply imaged
(Cordes, Pidwerbetsky, \& Lovelace~1987; \cite{cfl98}).  If stochastic
broadening dominates, the source structure should remain largely
unchanged except for an increase in its diameter.  The extent to which
one or the other of these mechanisms dominates may also provide clues
to the origin of these lenses.  To date, the only observational
constraints on the ESE mechanism---besides the light curves---have
been the lack of pulse broadening (as might be expected from a
stochastic broadening model) and a variation in the pulse time of
arrival (as would be expected from a refractive defocussing model)
during the ESEs toward PSR~B1937$+$21 (\cite{cblbadd93}; \cite{lrc98})
and PSR~J1643$-$1224 (\cite{mlc98}).  Of course, it need not be the
case that the ESE-like events observed toward these pulsars resulted
from the same kind of structure responsible for the ESEs toward
extragalactic sources.

This paper reports the first VLBI observations of a source
(1741$-$038, OT$-$068) while it was undergoing an \hbox{ESE}.  In
\S\ref{sec:observe} we describe the observations, in
\S\ref{sec:discuss} we discuss the implications of our observations,
and in \S\ref{sec:conclude} we present our conclusions and suggestions
for future work.

\section{Observations}\label{sec:observe}

Figure~\ref{fig:lightcurve} shows a portion of the 2.25~GHz light
curve of 1741$-$038 as obtained by the US Navy's extragalactic source
monitoring program at the Green Bank Interferometer
(\cite{fiedleretal87b}; \cite{wfjsfjmm91}; \cite{waltmanetal99}).
Clearly evident is an approximately 50\% decrease in the source's flux
density.  The minimum occurred on or near 1992 May~25
(JD~2448768.264), and the ESE is nearly symmetric about this epoch.
We show only a portion of the light curve in order to focus on the
\hbox{ESE}.  The complete GBI light curve of 1741$-$038, extending
from~1983 to~1994, has been published previously (\cite{cfl98}).

The epochs of observations are indicated in
Figure~\ref{fig:lightcurve}, and the observing log is given in
Table~1.  Also shown are the GBI flux density measurements
%Table~\ref{tab:log}.  Also shown are the GBI flux density measurements
closest in time to the VLBI observations.  While the GBI measurements
were not simultaneous with the VLBI observations, the largest elapsed
time between the GBI and VLBI observations was no more than 1~day.
All observations were conducted in VLBI Mk~II mode and recorded left
circular polarization only.  As part of this program, observations
were also obtained on 1993 June~2, well after the ESE concluded.

Shen et al.~(1997) conducted 6~cm VLBI observations in 1992~November.
Fey, Clegg, \& Fiedler~(1996a; see also Fey, Clegg, \& Fomalont~1996b; 
\cite{fc97}) have already presented images of 1741$-$038 at~6~cm
obtained on 1989 April~12 and at~3.6 and~13~cm obtained on 1994
July~8.  In all images, the source is dominated by a single compact
component, and there is little evidence of structural change within
the source during the interval 1991--1994.  These images have a higher
dynamic range than we could obtain from our 1993 June observations.
Consequently, we do not include the 1993 June observations here, but
instead use the images from the other groups to discuss source
characteristics outside the \hbox{ESE}.

% DUNCAN, R.A.; WHITE, G.L.; WARK, R.; REYNOLDS, J.E.; JAUNCEY, D.L.;
% NORRIS, R.P.; L. TAAFFE, SAVAGE, A., 
% ASTRON. SOC. OF AUSTRALIA. PROCEEDINGS V.10:4, P. 310, 1993
% has a visibility of 0.966 at 2.3 GHz on the Parkes-Tidbinbilla
% baseline corresponding to a source diameter of 15 mas
% 1988--1989

Nominal system temperatures for~1992 were obtained from the various
stations.  These were used for the initial amplitude calibration.  A
refined amplitude calibration was then determined in the following
manner.  The source 2121$+$053 (OX036) is a compact source used as a
fringe finder during the observing programs.  We fit a single,
circular Gaussian to observations of 2121$+$053 obtained in
1994~July,\footnote{
Observations obtained from the Radio Reference Frame Image Database
(RRFID),
$\langle$URL:http://www.usno.navy.mil/RRFID$\rangle$.
} finding a diameter of 0.9~mas.  With this source structure and the
source flux density measured from the GBI monitoring program, we then
fit the 2121$+$053 data at each epoch for a single, antenna-based
scaling factor.  This antenna-based scaling factor accounts for the
difference between the assumed and actual system temperatures.
Typical corrections were 10--20\%.

A crucial assumption, and a potential systematic error, of this method
is that the source diameter of 2121$+$053 remained essentially
constant over the interval 1992--1994.  VLBI images of 2121$+$053
obtained at epochs bracketing the time of our experiment show it to be
extremely compact (\cite{wcuaan92}; RRFID${}^{\thefootnote}$).  The
GBI monitoring program also shows no significant flux density changes,
such as a large increase in the flux density, indicative of structural
changes, such as the emergence of a new component.  As measured by the
GBI, the flux density of 2121$+$053 decreased by about 25\% during
1992, from about~2~Jy to~1.5~Jy, as part of a longer term decrease
following an outburst in 1988--1989.  During 1993 the flux density
stopped decreasing, remaining at about~1.5~Jy.  Superposed upon those
longer trends are shorter time scale variability; in the interval 1992
June--August, the flux density varied from near~2~Jy to as low as
1.3~Jy, then recovered to~1.7~Jy.

Fringe fitting at~13~cm was performed in a two steps.  The arrays used
consisted of a reasonably close cluster of antennas (mostly in the
southwest US) combined with a smaller number of far-flung antennas.
We fit first for the fringe delays and rates of the antennas
comprising the cluster, without solving for the delays and rates of
the distant antennas.  After applying these delay and rate solutions,
we fit for the delays and rates for the distant antennas only but used
all antennas in the fit.  This fringe fitting procedure increases the
probability that we will find fringes to all stations by focussing
first on the stations with the highest signal to noise and those that
should have the smallest rates and delays.

After fringe fitting, a series of first, phase-only and then,
amplitude-and-phase self-calibration iterations were used to account
for short time-scale fluctuations in antenna gain amplitudes and
phases.  Because of the extremely compact structure of 1741$-$038 seen
outside the ESE, we often used a point source as the model during
phase self-calibration.  Since the length of the GBI baseline (2.4~km)
is much shorter than the shortest baseline in our VLBI array (the
VLBA$\_$PT-VLBA$\_$LA baseline at 237~km), we used the GBI flux
densities as zero-spacing flux densities when imaging the source.

Unfortunately, because of either station or correlator problems during
the 1992 August~6 observations, only one hour of the 10-hour run had
more than three stations on source simultaneously.  With only three
stations available for most of the run, we can do little more than
phase-only self-calibration and fit simple models to the data.  We
shall therefore restrict the majority of our comments to the images
from the epochs 1992 June~8, June~20, and July~9.

The amount of time on-source ranged between $6^{\mathrm{h}}$ and
$10^{\mathrm{h}}$.  The resulting thermal noise limit on our maps is
therefore approximately 0.5~\mjybm.  The actual off-source rms noise
levels are in the range 1--3~\mjybm\ and are listed in
Table~1.
%Table~\ref{tab:log}.

Figures~\ref{fig:920608}--\ref{fig:920709} show the images for the
epochs 1992 June~8, June~20, and July~9.  We have fit one or more
Gaussian components to the $u$-$v$ data; the solutions to these fits
are in Table~\ref{tab:models}.  We discuss each epoch briefly and
separately.

\subsection{1992 June~8 (Fig.~\ref{fig:920608})}\label{sec:920608}

At this epoch the source is near the minimum flux density.  The source
structure at this epoch is consistent with that seen outside of the
ESE (1992 November, \cite{shenetal97}; 1989 April, 1994 July,
\cite{fcf96a}), namely a dominant compact component and a weaker
component to the south.  We had difficulty finding a model that fit
the data with a non-zero major axis for the secondary component,
whereas Fey et al.~(1996a) found a major axis of approximately 6~mas.
This discrepancy could result from two effects.  First, there could
have been modest evolution of the secondary component between~1992
and~1994.  Such evolution might also account for a portion of the
modest source variability seen in the GBI monitoring of 1741$-$038.
Second, our $u$-$v$ plane coverage is not as extensive as that of Fey
et al.~(1996a).  Consequently, our coverage may be sufficient to
indicate the component's presence without allowing us to fit a
detailed model to it.

\subsection{1992 June~20 (Fig.~\ref{fig:920620})}\label{sec:920620}

The source structure at this epoch consists of a single component.
The secondary component is not apparent.  Its apparent disappearance
is the combination of two effects.  First, the observation at this
epoch has only four stations, so that detailed structure is likely to
be lost.  Second, the secondary component has an inverted spectrum,
$\alpha \approx 2$ ($S_\nu \propto \nu^\alpha$, \cite{fcf96a}).
Extrapolating this spectrum to~18~cm, we expect that the flux
density of the secondary component will be 5--10~mJy, sufficiently
weak that we would not detect it.

%0.04~Jy @ 2.2~GHz  
%0.28~Jy @ 4.99~GHz
%0.75~Jy @ 8.55~GHz
% 2.2--4.99, \alpha = 2.4
% 4.99--8.55, \alpha = 1.8
% 2.2--8.55, \alpha = 2.2

In contrast to the 13~cm structure, the axial ratio at~18~cm is $b/a =
0.1$.  We attach little significance to this low value, however.
Because the array contains Hartebeesthoek, it is extended
significantly in the north-south direction.
%If we fit the data with a circular gaussian, the major axis increases
%to~0.3~mas.

\subsection{1992 July~9 (Fig.~\ref{fig:920709})}\label{sec:920709}

At this epoch the flux density of the source is starting to return to
its nominal value, but this epoch is prior to the peak flux density
during the egress of the \hbox{ESE}.  During the ingress and egress
from an ESE, when the flux density is above its nominal value,
ESE-induced changes in the source structure are most likely to be
visible (\cite{cfl98}).  Since we are unable to image the source at
its peak flux density (\S\ref{sec:920806}), this epoch represents our
best chance for seeing any ESE-induced changes in the source
structure.

The source structure is little changed from that outside the event.
The source continues to be dominated by a single compact component.
The secondary component is not apparent in the image at this epoch.
The image at this epoch has the highest off-source noise level of the
three epochs.  This high noise level is the result of numerous
intervals during the observing run in which fringes could not be
found.  Thus, this observation was more like a series of ``snapshots''
rather than a pointed observation.  The combination of high noise
level and poor $u$-$v$ coverage could contribute to difficulty in
detecting the secondary component.

Although limited $u$-$v$ coverage and the high noise level in this
image are at least partially responsible for the absence of the
secondary component, another possibility is that the ESE lens did not
cover the source fully.  Just such a possibility is indicated from
Clegg et al.'s~(1998) results.  They found that a lens comparable in
diameter to the compact component, 0.5~mas, produced the best match to
the light curve.  The relatively simple structure of 1741$-$038 does
not allow us to place any constraints on the axial ratio of the ESE
lens, though.

\subsection{1992 August~6}\label{sec:920806}

As noted above, for the majority of this epoch, only three stations
were observing the source.  Model fitting to the available data
indicates that the source continues to have a compact component.  The
model fitted to the visibility data is listed in
Table~\ref{tab:models}.

\section{Discussion}\label{sec:discuss}

In this section we use our VLBI images of 1741$-$038 to infer various
properties of the lens responsible for this ESE, and in particular,
the extent to which refractive and diffractive scattering were
important.  We begin by showing that the diffractive scattering, as
manifested by additional angular broadening of 1741$-$038 during the
event, did occur.  Clegg et al.~(1998) have modeled the 1741$-$038 as
due solely to refractive defocussing by an ionized cloud.  We discuss
the (limited) extent to which our observations can test the
predictions of their model.  We then turn to the question of the
mechanism by which ESEs can occur---refractive defocussing (e.g.,
\cite{rbc87}; \cite{cfl98}) or stochastic broadening (\cite{fdjws94}).
We conclude the section by considering what the shape of our images
implies about the shape of the lens and what the line of sight to
1741$-$038 implies about the genesis of the lens.

\subsection{Diffractive Properties of the 1741$-$038 ESE
	Lens}\label{sec:diffraction}

Our observable for studying the diffractive effects of this lens is
the angular broadening of the compact component of 1741$-$038.  Plasma
density fluctuations \emph{within} the lens will produce angular
broadening, in addition to any broadening resulting from density
fluctuations along the rest of the line of sight.  We shall show that
there is a measurable amount of angular broadening during the ESE,
first by comparing the angular diameter of 1741$-$038 at different
wavelengths during the ESE, then by comparing its angular diameter
during the ESE to that after the ESE at the same wavelength.  We shall 
then relate this additional angular broadening to the fluctuations
within the lens and show that the lens was probably quite turbulent
internally.

The density fluctuations responsible for interstellar scattering
(including angular broadening) are
typically parameterized by their spatial power spectrum as (Armstrong,
Rickett, \& Spangler~1995)
\begin{equation}
P_{\delta n_e} = \cnsq q^{-\alpha}.
\label{eqn:spectrum}
\end{equation}
The resulting scattering diameter, for a distant source and assuming
$\alpha = 11/3$ (\cite{r90}), is 
\begin{eqnarray}
\thd 
 & = & 128\,\mathrm{mas}\,\nu_{\mathrm{GHz}}^{-11/5}\mathrm{SM}^{3/5}, \nonumber \\
 & = & 1\farcs8\,\lambda_m^{11/5}\mathrm{SM}^{3/5},
\label{eqn:diameter}
\end{eqnarray}
where $\nu_{\mathrm{GHz}}$ is the observing frequency in GHz,
$\lambda_m$ is the observing wavelength in meters, and 
\begin{equation}
\mathrm{SM} = \int \cnsq(z)\,dz.
\label{eqn:sm}
\end{equation}
In the local ISM $\cnsq \sim 10^{-3.5}$~m${}^{-20/3}$ (\cite{ars95}),
and $\mathrm{SM} \sim 10^{-3.5}$~\smunits\ for a typical 1~kpc path
length through the local \hbox{ISM}.

During the ESE we find 13~cm diameters of~0.8 and~1~mas and an 18~cm
diameter of~1.6~mas (Table~\ref{tab:models}).  Taking the average of
the 13~cm diameters, we find the angular diameter scales as $\theta
\propto \lambda^{1.8}$.  We regard this wavelength scaling as
consistent with that expected for angular broadening, but not decisive
evidence in favor of it.  We have not attempted to correct for any
contribution by intrinsic structure, which should have a weaker
wavelength dependence than angular broadening.  Furthermore, some
deviation from a strict $\lambda^{2.2}$ dependence might be expected
because we are comparing angular diameters determined at different
epochs during a time-dependent event.  Indeed, the angular diameter
at~13~cm does change in exactly the manner expected if the lens is
centrally condensed; near the flux density minimum (1992 June~6 epoch)
when the path length through a centrally condensed lens would be near
the maximum, the angular diameter is slightly larger (1~mas) as
compared to the angular diameter (0.8~mas) near the end of the ESE
(1992 July~9) when the path length through the lens would be shorter.
However, our determination of the $\lambda^{1.8}$ scaling does depend
crucially upon the 18~cm diameter, and, in turn, on the limited
$u$-$v$ coverage for this observation.  If we fit a circular Gaussian
instead of an elliptical Gaussian to the 18~cm visibility data, we
find an angular diameter of~0.3~mas, implying a $\lambda^{-3.4}$
dependence.  Hence, we regard the comparison of angular diameters
during the ESE as suggestive, but not compelling, evidence, for
increased angular broadening during the \hbox{ESE}.

We find a more compelling demonstration of an increase in the angular
diameter from comparing the angular diameter of 1741$-$038 during and
after the \hbox{ESE}.  During the ESE the 13~cm diameter of 1741$-$038
was $0.9 \pm 0.1$~mas.  In a series of subsequent observations, during
1994--1997 (\cite{fcf96a}; \cite{fcf96b}; \cite{fc97}; Fey,
unpublished data), acquired while the source was not undergoing an
ESE, the angular diameter of the compact component has been measured
to be 0.5 to~0.75~mas, with a mean of $0.63 \pm 0.04$~mas.  The quoted
uncertainty in the mean diameters are statistical.  Systematic
effects, namely the length of the longest baseline in the VLBI array
used, probably contribute to an overestimation of the diameter.  The
same systematic effects appear to contribute to our fits also being
overestimates of the actual diameter.  Time-dependent changes may also
affect the diameters during the \hbox{ESE}.

We find the excess angular broadening due to the ESE lens by
subtracting in quadrature the diameters during and after the \hbox{ESE}.
Using the diameters determined above, namely $\theta_{\mathrm{in}} =
0.9$~mas and $\theta_{\mathrm{out}} = 0.63$~mas, we find $\delta\thd
\equiv \sqrt{\theta_{\mathrm{in}}^2 - \theta_{\mathrm{out}}^2}
\lesssim 0.7$~mas.  We treat this value as an upper limit based on our
assessment of the systematic uncertainties in the measured diameters.
However, our longest baseline is often significantly longer than that
used in determining the diameters after the ESE, so our measurement of
an increase in the diameter of the source is robust.

The excess angular broadening of the lens arises from an additional
\cnsq\ in the line of sight to 1741$-$038 during the \hbox{ESE}.  We
solve for the level of scattering within the lens,
$\mathrm{SM}_{\mathrm{lens}}$, in the following manner.  The
scattering within the lens is given by 
\begin{equation}
\mathrm{SM}_{\mathrm{lens}} \equiv \mathrm{SM}_{\mathrm{in}}
 - \mathrm{SM}_{\mathrm{out}},
\label{eqn:smdiff}
\end{equation}
where $\mathrm{SM}_{\mathrm{in}}$ and $\mathrm{SM}_{\mathrm{out}}$ are 
the scattering measures seen inside and outside the ESE, respectively.

We have estimated $\mathrm{SM}_{\mathrm{out}}$ from RRFID observations 
(\cite{fcf96a}; \cite{fcf96b}; \cite{fc97}; Fey, unpublished data).
The RRFID observations acquire 3.6 and~13~cm data simultaneously.  We
have fit the measured 3.6 and~13~cm diameters to
\begin{equation}
\theta^2(\lambda) = \theta_{d,1}^2\lambda_m^{4.4}
 + \theta_{I,1}^2\lambda_m^2,
\label{eqn:adfit}
\end{equation}
assuming that the scattering and intrinsic diameters add in quadrature
and that the intrinsic diameter scales as $\lambda^1$, as is
appropriate for a synchrotron self-absorbed component (\cite{ko88}).
Here $\theta_{d,1}$ and $\theta_{I,1}$ are the scattering and
intrinsic diameters, respectively, at the fiducial wavelength of~1~m.
The spectral index of 1741$-$038 is $\alpha \approx 0.17$ ($S
\propto \nu^\alpha$) around the time of the ESE, but
excluding the ESE itself, and $\alpha \approx 0.37$ near the time
of the RRFID observations, in both cases consistent with the
assumption of self-absorption.  We then use
equation~(\ref{eqn:diameter}) to solve for
$\mathrm{SM}_{\mathrm{out}}$.  We find $\mathrm{SM}_{\mathrm{out}} =
10^{-3}$~\smunits.

This value of $\mathrm{SM}_{\mathrm{out}}$ is consistent with the
available constraints on the scattering diameter of 1741$-$038 from
low frequencies.  Using interplanetary scintillation observations
at~92~cm, Vijayanarasimha et al.~(1985) determined that 1741$-$038
must have a component with a diameter of~100~mas.  Assuming that
interstellar scattering dominates at this wavelength, we find a
scattering measure $\mathrm{SM}_{\mathrm{out}} \le 10^{-2}$~\smunits.
We treat this value as an upper limit because IPS observations do not
give detailed information on the source structure.

We estimate $\mathrm{SM}_{\mathrm{in}}$ from the difference in the
diameter of 1741$-$038 inside and outside of the ESE,
\begin{equation}
(\delta\thd)^2
 = \theta_{\mathrm{in}}^2 - \theta_{\mathrm{out}}^2
 = (1\farcs8)^2\lambda_m^{22/5}\left(\mathrm{SM}_{\mathrm{in}}^{6/5} - \mathrm{SM}_{\mathrm{out}}^{6/5}\right).
\label{eqn:addiff}
\end{equation}
We have already constrained $\delta\thd \le 0.7$~mas.  We find
$\mathrm{SM}_{\mathrm{in}} \sim 10^{-2.4}$~\smunits, a value of SM
that is somewhat larger than the typical SM through the local
\hbox{ISM}.

Thus, the ESE lens toward 1741$-$038 produced an additional angular
broadening of approximately 0.7~mas, and the lens itself had a
scattering measure of $\mathrm{SM}_{\mathrm{lens}} =
10^{-2.5}$~\smunits.

In order to produce a significant flux density suppression during the
ESE, from stochastic broadening alone, the angular diameter of the
lens must be comparable to that of the source (\cite{fdjws94}).  We
assume that the lens diameter was $a \sim 1$~mas (\S\ref{sec:esemech})
or $a \sim 0.1\,\mathrm{AU}(D/0.1\,\mathrm{kpc})$, where $D$ is the
distance to the lens.  Because the extent of the lens along the line
of sight may be (considerably) different than its transverse size, we
take $\mathrm{SM}_{\mathrm{lens}} = C_{n,\mathrm{lens}}^2\eta a$,
where the factor $\eta$ is the ratio of the lens' extent along the
line of sight to its transverse extent.  We find
$C_{n,\mathrm{lens}}^2 \sim
10^7\eta^{-1}\,\mathrm{m}{}^{-20/3}(D/0.1\,\mathrm{kpc})^{-1}$.  For
comparison, Hjellming \& Narayan~(1986) estimated that $\cnsq \gtrsim
10^{-1.5}$~m${}^{-20/3}$ for this line of sight using the refractive
scintillation of this source outside of the \hbox{ESE}.

%Clegg et al.~(1998) were able to infer a transverse lens size for the
%1741$-$038 event by fitting the 2.2~GHz light curve and assuming that
%the ESE lens had a gaussian electron density profile.  They inferred a
%transverse lens size of $a =
%0.065\,\mathrm{AU}(D/0.13\,\mathrm{kpc})$, where $D$ is the distance
%to the lens.  Because the extent of the lens along the line of sight
%may be (considerably) different than its transverse size, we take
%$\mathrm{SM}_{\mathrm{lens}} = C_{n,\mathrm{lens}}^2\eta a$, where the
%factor $\eta$ is the ratio of the lens' extent along the line of sight
%to its transverse extent.  We find $C_{n,\mathrm{lens}}^2 \sim
%10^7\eta^{-1}\,\mathrm{m}{}^{-20/3}(D/0.13\,\mathrm{kpc})^{-1}$.  For
%comparison, Hjellming \& Narayan~(1986) estimated that $\cnsq \gtrsim
%10^{-1.5}$~m${}^{-20/3}$ for this line of sight using the refractive
%scintillation of this source outside of the \hbox{ESE}.

One of the key features of ESE-like events observed toward pulsars is
that the pulse width does not increase (\cite{cblbadd93};
\cite{lrc98}; \cite{mlc98}).  Using our estimate of
$\mathrm{SM}_{\mathrm{lens}}$, we can predict how much pulse
broadening, another diffractive effect, would be produced from a lens
comparable to that which passed in front of 1741$-$038.  The amount of
pulse broadening is (\cite{tc93})
\begin{equation}
\tau_d
 \le 1.1\,\mathrm{ms}\,D_{\mathrm{kpc}}(\mathrm{SM})^{6/5}\nu_{\mathrm{GHz}}^{-22/5}.
\label{eqn:taud}
\end{equation}
We can calculate only an upper limit because the amount of pulse
broadening depends upon the location of the lens along the line of
sight (\cite{cr98}); the maximum occurs when the lens is midway
between the observer and pulsar.  We predict that ESE lenses typical
of the one that passed in front of 1741$-$038 will increase the pulse
broadening of a background pulsar by only $1.1D_{\mathrm{kpc}}$~$\mu$s
at~1~GHz.  This small amount is consistent with the lack of broadening
seen toward the millisecond pulsars PSR~B1937$+$21 and
PSR~J1643$-$1224.

Our analysis has assumed that scattering within this lens can be
described in terms of a power-law spectrum of density fluctuations
(eqn.~[\ref{eqn:spectrum}]).  While this may be true within the lens
(\S\ref{sec:spectrum}), the lens itself cannot be formed by the same
processes that give rise to the spectrum of density fluctuations in
the local ISM (\cite{ars95}).  First, a medium pervaded by density
fluctuations on AU scales should produce a correlation between flux
density and angular diameter (\cite[their Figs.~1 and~4]{bn85}).  In
contrast, we observe an \emph{anti-correlation}, with the angular
diameter increasing as the flux density decreases.  Second, if the
values we have used for~$a$ and~$D$ are not severe underestimates, the
large value of $C_{n,\mathrm{lens}}^2$ we infer is significantly
larger than the value in the local \hbox{ISM}.  Even allowing for a
structure extremely elongated along the line of sight, $\eta > 100$,
$C_{n,\mathrm{lens}}$ remains orders of magnitude above the value in
the local \hbox{ISM}.  We regard the large value of
$C_{n,\mathrm{lens}}^2$ as an indication that the genesis of an ESE
lens requires an energetic process.

\subsubsection{The Electron Density Power Spectrum within an ESE
	Lens}\label{sec:spectrum}

The measured visibility on an interferometer baseline~$b$ is
\begin{equation}
V(b) = e^{-D_\phi(b)/2},
\label{eqn:visibility}
\end{equation}
for a point source seen through a region of density fluctuations with
a spatial power spectrum given by equation~(\ref{eqn:spectrum}).  The
phase structure function~$D_\phi(b)$ is a measure of the fluctuations
induced in the wavefront's phase as it propagates through the
scattering medium and is given by (e.g., \cite{cl91})
\begin{equation}
D_\phi(b) \propto \thd^\beta b^\beta.
\label{eqn:dphi}
\end{equation}
Here $\thd$ is the diffractive scattering angle, and $\beta \equiv \alpha -
2$.

There are a number of lines of sight that suggest $\alpha \approx
11/3$, the Kolmogorov value (\cite{r90}).  There are also some lines
of sight that suggest a significantly larger value, $\alpha > 4$.  The
diameters tabulated in Table~\ref{tab:models} were found by fitting a
gaussian to the visibilities, i.e., assuming $\beta = 2$.  We now
relax that requirement.

For the three epochs for which we were able to produce images, we fit
the visibility data with a model of the form of
equation~(\ref{eqn:visibility}).  For all three epochs we were unable
to place any meaningful constraints on $\beta$.  Allowed values of
$\beta$, at all three epochs, were 1--2.2.  Though the range is nearly
centered on the Kolmogorov value of~1.67, we cannot exclude $\beta \ge
2$.

\subsection{Refractive Properties of the 1741$-$038 ESE
	Lens}\label{sec:refraction}

Refractive effects expected from ESE lenses include substructure
within the source, angular position wander, and multiple imaging,
though the extent to which any of these occur depends upon the
strength of refraction within the lens.  There is no indication of
refractively-induced substructure in the source such as might be
produced if the lens had substructure within it.  During the event,
the source consisted of a compact component, with a weak secondary
component detectable to the south in at least one epoch.  This
structure is essentially identical to that seen after the event
(\cite{shenetal97}; \cite{fcf96a}).  Below, we quantify the
possibility that the shape of the lens has altered the shape of the
source (\S\ref{sec:anisotropic}).

One of the key predictions of a refractive model for ESEs is that an
ESE should produce angular position wander of the background source.
Clegg et al.~(1998) predicted that during the 1741$-$038 \hbox{ESE},
the angular position of the source wandered by~0.4~mas at~13~cm and
by~0.8~mas at~18~cm.  The observations reported here were not
phase-referenced, and the self-calibration we performed erased all
absolute position information.  Furthermore, the proximity of the two
intrinsic components of 1741$-$038, the sparse visibility data, and
limited dynamic range (particularly of the 1992 July~9 observations)
give us little confidence of detecting relative position shifts
between the two components, such as would occur if the lens covered
only one component at a time.  Consequently, even if the lens covered
only the brighter component, it would be difficult to determine the
relative separation, with any degree of confidence, between the two
components in either the image or visibility domains.  We are thus
unable to test the prediction of ESE-induced angular position wander.

A second prediction is that an ESE can produce multiple imaging of the
background source.  In the case of this ESE, any secondary image(s)
must have been either extremely faint or only slightly offset with
respect to the primary image.  within the 10--15\% uncertainties of
the GBI-measured flux densities (\cite{fiedleretal87b}) and those in
our amplitude calibration, which we estimate to be at least 10\%, our
models account for all of the flux density measured by the \hbox{GBI}.
A bright secondary image(s) could have been present only if the
multiple images nearly overlapped so that no significant anisotropy
was produced in our VLBI images (Table~\ref{tab:models}).  Clegg et
al.~(1998) predicted that this ESE was not strong enough to form
caustics and produce multiple imaging.

\subsection{The ESE Mechanism}\label{sec:esemech}

As noted in \S\ref{sec:intro}, two general classes of models have been
advanced to explain how ESEs occur.  In the refractive defocussing
model (\cite{rbc87}; \cite{cfl98}; \cite{ww98}; hereinafter the RD
model) the decrease in the source's flux density during the event
occurs because of refractive defocussing of the incident rays on the
lens.  In the stochastic broadening model (\cite{fdjws94}; hereinafter
the SB model) the flux density decrease occurs because small-scale
inhomogeneities in the lens scatter the incident wave front.  While
most recent work has focussed on the RD model---particularly the work
of Clegg et al.~(1998) who compared quantitatively the observed light
curve and that predicted from an RD model---both models can reproduce
the generic features of an ESE light curve, and the only other
observational data with which to compare the models have been the
pulsar timing programs described in \S\ref{sec:intro}.

A key prediction of the refractive model is that the source's flux
density and angular diameter should be highly correlated.  In contrast
we observe an \emph{anti-correlation}, more consistent with that
expected from the SB model.  Fey et al.~(1996b) discussed
qualitatively how the SB model could produce the 1741$-$038 ESE, but
did not compare quantitatively the observed light curve and the SB
model.  Consequently, we shall re-visit the question of the mechanism
by which ESEs are produced and whether the SB model can account for
this \hbox{ESE}.

We shall use the SB model developed by Fiedler et al.~(1994,
Appendix~A).  This model describes the flux density of a source during
an ESE as $\hat F(t; F_0, \mu, \theta_I, \theta_\ell, \theta_b)$.
Here $F_0$ is the source's nominal flux density outside the lens,
$\mu$ is the proper motion of the lens across the lie of sight,
$\theta_I$ is the intrinsic (FWHM) angular diameter of the background
source, $\theta_\ell$ is the apparent angular width of the lens, and
radiation incident on the lens is scatter broadened by $\theta_b$
(FWHM).

We used a grid-search method to search the available parameter space,
evaluating the goodness of fit by the $\chi^2$ statistic.  Based on
the measurements of 1741$-$038 outside the ESE and our determination
of the additional angular broadening during the event
(\S\ref{sec:diffraction}), we began by holding $\theta_I$ and
$\theta_b$ fixed at $\theta_I = 0.5$~mas and $\theta_b = 0.7$~mas.  We
thus fit for the parameters $F_0$, $\mu$, and~$\theta_\ell$.

We were unable to find reasonable agreement.  In particular the
best-fit model had a flux density minimum that was approximately 85\%
of the nominal flux density as opposed to the 50\% minimum that was
observed.  Furthermore the shape of the modeled ESE light curve is
that of a flat-bottomed minimum rather than the rounded minimum
observed.  We then removed the constraints on~$\theta_I$
and~$\theta_b$ separately.  If we fit for $\theta_I$, $F_0$, $\mu$,
and~$\theta_\ell$ while holding $\theta_b$ fixed, the fit agreement
improves slightly.  However, the depth of the modeled flux density
minimum continues to be insufficient to match the observed depth, and
the minimum is flat-bottomed as opposed to the observed rounded
minimum.

If we fit for $\theta_b$, $F_0$, $\mu$, and~$\theta_\ell$ while
holding $\theta_I$ fixed, we find quite reasonable agreement with both
the depth and shape of the minimum being reproduced.
Table~\ref{tab:sbmodel} lists our best-fitting parameters.  The
significant result of our fit is that the broadening diameter required
to reproduce the observed light curve is much larger than what we
infer from our measurements.  The SB model alone requires a broadening
angle $\theta_b = 2$~mas, while our measurements suggest that only an
additional 0.7~mas of broadening occurred during the event.

Moreover, the value of $\theta_b$ found is probably a lower limit.
The model of Fiedler et al.~(1994) assumes that $\theta_b$ is constant
across the width of the lens.  If the strength of broadening varies
across the lens, being stronger in the center and weaker around the
edges, an even larger value of $\theta_b$ would be required to obtain
the same decrease in flux density during the event.

We therefore conclude that the SB model alone cannot explain both the
observed light curve and amount of angular broadening.  Given the good
agreement that Clegg et al.~(1998) found using the RD model, we
consider it likely that both refractive defocussing and stochastic
broadening are occurring.

\subsection{Image Anisotropy}\label{sec:anisotropic}

The images shown in Figures~\ref{fig:920608}--\ref{fig:920709} display
varying degrees of anisotropy.  We can identify six causes of image
anisotropy (\cite{cpl87}; \cite{c90}; see also \cite{sc88}):
(1)~incomplete $u$-$v$ coverage; (2)~intrinsic source structure;
(3)~anisotropic diffractive scattering; (4)~anisotropic refractive
focussing; (5)~unresolved multiple images; and (6)~a spatially-limited
scattering volume.

The first cause---incomplete $u$-$v$ coverage---is almost certainly
the explanation for the image of 1992 June~20.  The array was
elongated significantly in the north-south direction because it
included Hartebeestock.  The $u$-$v$ coverage was considerably more
uniform for the other two epochs, though, and incomplete $u$-$v$
coverage is less likely to be the cause of the image anisotropy at
these epochs.

While we cannot rule out the possibility of one or more of the
remaining causes contributing to the source anisotropy, we can set
limits on the shape of the lens by assuming all of the image
anisotropy is due to refractive focussing.  The electron density
profile across the lens produces a refractive gain~$G$.  If the
density profile in two orthogonal directions differs, the source will
have an axial ratio (\cite{cpl87}; \cite{sc88})
\begin{equation}
\frac{b}{a} = \frac{G_b}{G_a}.
\label{eqn:axialratio}
\end{equation}

The axial ratios we measure during the ESE are $b/a > 0.8$.  These
axial ratios are much closer to unity than would be expected if the
lens had an axial ratio~$\eta \sim 100$ (e.g., \cite{rbc87}), and its
long axis was in or near the plane of the sky.  We conclude that the
refractive strength of the ESE lens was not considerably different in
different directions on the plane of the sky.

Romani et al.~(1987) have speculated that ESE lenses are filamentary
structures.  Filamentary structures have the desirable property of
reducing the degree to which the ESE lenses are overpressured with
respect to the nominal interstellar pressure.  As we discuss above,
filamentary structures would also reduce the inferred level of
turbulence within the lenses.  Filamentary ionized structures would
also form naturally in a magnetized medium.  If ESE lenses are
filamentary structures, the nearly isotropic image shapes that we
observe indicate that the lenses are extended \emph{along} the line of
sight.  Lestrade et al.~(1998) have suggested that ESEs toward pulsars
occur only when filamentary or sheet-like structures are favorably
oriented along the line of sight.  Our VLBI images suggest that a
similar situation is necessary for ESEs toward extragalactic sources.

\subsection{Environments of ESE Lenses}\label{sec:environs}

Fiedler et al.~(1994b) showed that the lines of sight to a small
number of sources that have undergone an ESE display structures
suggestive of a turbulent origin.  For instance, the line of sight to
0954$+$658 passes near the edge of radio Loop~III, with filamentary
structures seen in 100$\mu$ emission, and the line of sight to
2352$+$495 (DA~611) passes near RAFGL~5797S, an infrared source with a
cometary morphology.  Fielder et al.~(1994a) argued that the
distribution of ESE sources near the edges of these radio loops was
not accidental, but was indicative of a connection between ESEs and
sites of interstellar turbulence such as old supernova remnants.
Supernova remnants could also provide a high-pressure environment in
which the ESE lenses could survive (\cite{rbc87}).  Walker \&
Wardle~(1998) have since suggested that ESEs arise from dense
molecular clouds in the Galaxy's halo.

The line of sight to 1741$-$038 displays similar evidence of strong
gradients and turbulent-like structures.
Figure~\ref{fig:env}\textit{a} shows the \ion{H}{1} column density
toward the line of sight of 1741$-$038 and 1749$+$096 (4C~09.57),
another source observed to have undergone an \hbox{ESE}.  The
\ion{H}{1} column density displays a strong gradient with Galactic
latitude, changing by a factor of nearly $10^2$ over 10\arcdeg\ in
latitude.  Furthermore, the two spurs of \ion{H}{1} emission extending
to higher latitudes in Figure~\ref{fig:env}\textit{a} are part of
radio Loop~I, a structure Fiedler et al.~(1994a) have already
suggested is responsible for the ESE toward 1749$+$096.  Like
1749$+$096, 1741$-$038 is close to, perhaps within, a portion of
\hbox{Loop~I}.  Figure~\ref{fig:env}\textit{b} shows the 100$\mu$
emission toward 1741$-$038.  While not as dramatic as the structures
seen along the lines of sight toward 0954$+$658 and 2352$+$495,
1741$-$038 does appear near a local minimum in the 100$\mu$ emission.
The appearance of the emission is suggestive of a process which has
excavated a cavity in the \hbox{ISM}.  Ionized gas, with significant
density enhancements, could result within the cavity or on its edges.

We cannot conclude, from these \ion{H}{1} and 100$\mu$ images alone,
that ESEs are produced at sites of interstellar turbulence, but these
images do add to the existing circumstantial evidence suggesting that
this is the case.

It might also be possible to place further constraints on the distance
to the material responsible for ESEs in this direction.  The pulsar
PSR~J1743$-$0337 (PSR~B1740$-$03) is located only 18\arcmin\ away from 
1741$-$038.  It has a dispersion measure of 35~pc~cm${}^{-3}$,
corresponding to a distance estimate of~1.8~kpc (\cite{tc93}).  If
this pulsar showed enhanced scattering or refractive events, like ESEs 
or fringing in a dynamic spectrum, that would be a strong indication
that at least a portion of the scattering in this direction occurs
because of material closer than 1.8~kpc.

\section{Conclusions}\label{sec:conclude}

We have presented the first VLBI images of a source, 1741$-$038,
obtained at multiple epochs as the source underwent an extreme
scattering event.

We have used these images to assess two models for the origin of
ESEs---a refractive defocussing model and a stochastic broadening
model.  The source structure is dominated by a compact component and
is essentially unchanged during the event as compared to the structure 
after the event.  The only change is a slight increase in the diameter 
of the source (by~0.7~mas), an increase we attribute to additional
angular broadening within the lens.  This additional angular
broadening is consistent with that expected from a stochastic
broadening model but is \emph{not} consistent with that expected from
a purely refractive defocussing model.  Specifically, a refractive
defocussing model predicts a correlation between the flux density and
angular diameter of the source.  However, attempts to reproduce the
ESE light curve of 1741$-$038 by a purely SB model require a larger
increase in the source's diameter (2~mas) than is observed (0.7~mas).
We cannot test other predictions of the RD model.  The refractive defocussing model predicts angular
position wander of the source, but our observations were not sensitive
to absolute angular
position shifts.  We also see no evidence of strong multiple imaging,
but a quantitative comparison of the ESE light curve and an RD model
predicts that the refractive strength of the lens was not sufficient
to produce multiple imaging (\cite{cfl98}).  We conclude that the
1741$-$038 ESE involved both SB and RD processes.  

The angular diameter of 1741$-$038 increased by about 0.7~mas during
the \hbox{ESE}.  The amount of angular broadening contributed by the
lens implies that the interiors of lenses are highly turbulent with
levels of scattering orders of magnitude higher than that seen in the
local \hbox{ISM}.  The inferred level of the electron density power
spectrum, as parameterized by the coefficient~\cnsq, is
$C_{n,\mathrm{lens}}^2 \sim
10^7\eta^{-1}\,\mathrm{m}{}^{-20/3}(D/0.1\,\mathrm{kpc})^{-1}$.  A
filamentary lens, with $\eta > 1$, would decrease the required \cnsq.

The observed visibility data are consistent with the interior of the
lens having a power-law density power spectrum, with a power spectral
index similar to that seen in the local ISM, though we cannot rule out
a ``steep'' density spectrum ($\beta > 4$).  The lens itself could not 
have arisen from the density fluctuations in the local ISM, however.
The value of $C_{n,\mathrm{lens}}^2$ is well in excess of the local
value, and evenly distributed density fluctuations would give rise 
to a flux density-angular diameter correlation in contrast to the
observed anti-correlation.

If ESE lenses are filamentary, as has been suggested to reduce their
overpressure relative to the ambient medium, ESEs must occur only when
the filamentary structures are seen nearly lengthwise.  A filamentary
lens seen transverse to its long axis would produce different
refractive gains along and across the lens, resulting in image
anisotropy.  Our images display little anisotropy.  

The line of sight toward 1741$-$038 shows a strong gradient in the
neutral hydrogen density, and the source lies close to or within radio
Loop~I, with small-scale, ``cavity''-like structure seen at~100$\mu$.
Such interstellar structures along the line of sight to 1741$-$038 are
similar to that seen toward some of the other sources for which ESEs
have been observed.  This line of sight is thus consistent with ESE
lenses originating from energetic turbulent processes in the
\hbox{ISM}.

Future observations of a source undergoing an \hbox{ESE} will be
enhanced by the existence of dedicated VLBI arrays such as the Very
Long Baseline Array.  A key prediction of the refractive model---one
that we have been unable to test---is the existence of angular
wandering.  Future observations should also have a more extensive and
uniform $u$-$v$ plane coverage, making the imaging process easier.
Modern, frequency-agile receivers also allow for the possibility of
simultaneous or nearly simultaneous images at multiple frequencies.
Observations at multiple frequencies should include imaging the source
in the \ion{H}{1} line to search for the existence of neutral
structures related to the ionized structures responsible for ESEs.
The major impediment to a set of such observations is the lack of a
existing monitoring program that could find additional ESEs.

\acknowledgements
We thank S.~Spangler for helpful discussions regarding interstellar
refraction, and K.~Desai for his image fitting software.  This
research made use of NASA's Astrophysics Data System Abstract Service;
the SIMBAD database, operated at the CDS, Strasbourg, France; and
NASA's \textit{SkyView} facility\footnote{
$\langle$URL:http://skyview.gsfc.nasa.gov/$\rangle$
} located at NASA Goddard Space Flight Center.  A portion of this work
was performed while TJWL held a National Research Council-NRL Research
Associateship.  Basic research in radio astronomy at the NRL is
supported by the Office of Naval Research.

\clearpage

\begin{figure}
\caption[]{The extreme scattering event of 1741$-$038.  The dots show
the 2.2~GHz (13~cm) flux density as measured by the Green Bank Interferometer.
The vertical lines mark the epochs at which the VLBI observations
reported in this paper were obtained.}
\label{fig:lightcurve}
\end{figure}

\begin{figure}
\caption[]{The epoch 1992 June~8 at the wavelength 13~cm.
(\textit{a})~The image.  The off-source noise level is
1~\mjybm, and contours are 1~\mjybm\ $\times$ $-3$, 3, 5, 10, 20,~\ldots.
The beam is shown in the lower left.
(\textit{b})~The visibility data as a function of projected baseline.
A model of two components is shown superposed for reference.  One
component is a \emph{circular} gaussian with an amplitude of~1.15~Jy
and a diameter equivalent to that implied by the model fit from
Table~\ref{tab:models}, namely 1~mas.  The second component is a delta
function with an amplitude of~0.01~Jy located 9.7~mas to the south of
the first component.}
\label{fig:920608}
\end{figure}

\begin{figure}
\caption[]{The epoch 1992 June~20 at the wavelength 18~cm.  
(\textit{a})~The image.  The off-source noise level is 1.6~\mjybm, and
contours are 1.6~\mjybm\ $\times$ $-2$, 3, 5, 10, 20,~\ldots.  The beam is shown in the lower left.
(\textit{b})~The visibility data as a function of projected baseline.
A model of a single, \emph{circular} gaussian component is shown
superposed for reference.  The gaussian component has an amplitude
of~0.95~Jy and a diameter equivalent to that implied by the model fit
from Table~\ref{tab:models}, namely 0.61~mas.}
\label{fig:920620}
\end{figure}

\begin{figure}
\caption[]{The epoch 1992 July~9 at the wavelength 13~cm.  
(\textit{a})~The image.  The off-source noise level is 3.3~\mjybm, and
contours are 3.3~\mjybm\ $\times$ $-3$, 3, 5, 10, 20,~\ldots.  The beam is
shown in the lower left.
(\textit{b})~The visibility data as a function of projected baseline.
A model of a single, \emph{circular} gaussian component is shown
superposed for reference.  The gaussian component has an amplitude
of~2.16~Jy and a diameter equivalent to that implied by the model fit
from Table~\ref{tab:models}, namely 0.93~mas.}
\label{fig:920709}
\end{figure}

\begin{figure}
\caption[]{The environs of 1741$-$038.  \textit{(a)}  The gray scale
shows the column density of \ion{H}{1} (\cite{dl90}), ranging between
$10^{20.8}$~cm${}^{-2}$ (white) and $10^{22.1}$~cm${}^{-2}$ (black).
The positions of 1741$-$038 and 1749$+$096, another source observed to
have undergone an \hbox{ESE}, are marked.
\textit{(b)} The gray scale shows the 100$\mu$ emission, as derived from
\hbox{IRAS} observations, ranging between 17~MJy~sr${}^{-1}$ (white)
and 37~MJy~sr${}^{-1}$ (black).  The position of 1741$-$038 is marked.}
\label{fig:env}
\end{figure}

\clearpage

\begin{deluxetable}{lccccccc}
\tablenum{2}
\tablecaption{Source Models\tablenotemark{a}\label{tab:models}}

\tablehead{\colhead{Epoch} 
	& \colhead{$\lambda$} & \colhead{$S$}
	& \colhead{$r$} & \colhead{$\psi$} 
	& \colhead{$a$} & \colhead{$b/a$} & \colhead{$\phi$} \\
	& \colhead{(cm)} & \colhead{(Jy)}
	& \colhead{(mas)} & \colhead{(\arcdeg)}
	& \colhead{(mas)} & \colhead{(\arcdeg)}}

\startdata
1992 June~8   & 13 & 1.15 & 0.0 & \phn\phn0 & 1.0 & 0.83          &    -9.5\nl
		   & 0.01 & 9.7 &       179 & 0.0 & 1\phd\phn\phn & \nodata\nl

1992 June~20  & 18 & 0.95 & 0.0 & \phn\phn0 & 1.7 & 0.1           & 69\nl
%1992 June~20  & 18 & 0.95 & 0.0 & \phn\phn0 & 0.3 & 1.0           & 0\nl

1992 July~9   & 13 & 2.16 & 0.0 & \phn\phn0 & 0.98 & 0.91          & \nodata\nl

1992 August~6 & 13 & 2.52 & 0.0 & \phn\phn0 & 1.5 & 1            & \nodata\nl
\enddata
\tablenotetext{a}{Source models consist of gaussians of flux
density~$S$, major axis~$a$, and axial ratio~$b/a$ at position
angle~$\phi$ located a distance~$r$ from the phase center at a
position angle~$\psi$.  The stronger component was always assumed to
be at the phase center.}

\end{deluxetable}

\begin{deluxetable}{cc}
\tablenum{3}
\tablecaption{Parameters of the Best-Fit Stochastic Broadening 
	Model\label{tab:sbmodel}}
\tablehead{\colhead{Parameter} & \colhead{Value}}
\startdata
$\theta_I$\tablenotemark{a} & 0.5~mas \nl
$\theta_b$  & 2~mas \nl
$F_0$       & 1.9~Jy \nl
$\theta_\ell$ & 1~mas \nl
$\mu$       & 14~$\mu$as~d${}^{-1}$ \nl
\nl
$\chi^2$    & 16 \nl
\enddata
\tablenotetext{a}{This parameter held fixed.}
\end{deluxetable}

\end{document}